\def\be{\begin{equation}}
\def\ee{\end{equation}}
\def\bea{\begin{eqnarray}}
\def\eea{\end{eqnarray}}
\def\ba{\begin{align*}}
\def\ea{\end{align*}}

\documentclass[twocolumn,prl,amsmath,amssymb,apsrev,groupedaddress,superscriptaddress,showpacs,showkeys]{revtex4}
\usepackage{amsfonts,amssymb,latexsym,amsmath,epsfig,bm,times}
\usepackage{color}
\usepackage{bm}
\usepackage{graphicx}
\begin{document}
\title{Origin of Lattice Spin in Graphitic Systems}
\author{Kumar Abhinav}
\email{kumarabhinav@iiserkol.ac.in}
\affiliation {Indian Institute of Science Education and Research Kolkata, Mohanpur-741246, West Bengal, India}
\author{Prasanta K. Panigrahi}
\email{pprasanta@iiserkol.ac.in}
\affiliation {Indian Institute of Science Education and Research Kolkata, Mohanpur-741246, West Bengal, India}
\date{\today}
\begin{abstract}
Lattice spin, in planar condensed matter system with emergent Dirac dispersion, is 
shown to emerge from the inherent $SU(2)$ symmetry, arising through 
Schwinger's angular momentum construction from anti-commuting Heisenberg operators of the sub-lattices. 
The presence of a mass term in the emergent Dirac dispersion is essential for the existence of this spin. The usual hopping term,
that entangles the two sub-lattices, leads to the orbital counterpart. Relative sub-lattice displacements, that couple to
the effective Dirac fermions like $U(1)$ gauge fields, do not effect the lattice spin.  
\end{abstract}
\keywords{Graphene, Lattice Spin, $SU(2)$ Symmetry}
\pacs{73.22.Pr,75.10.Jm,03.65.Fd}
\maketitle

By construction, Dirac fermions, characterized by the Hamiltonian,

\be
{\cal H}=c{\boldsymbol\alpha}\cdot{\bf p}+\beta mc^2,\label{01}
\ee
in the momentum space, do not commute with the orbital angular momentum (OAM) operator,
${\bf L}={\bf r}\times{\bf p}$. This is the origin of spin angular momentum (SAM) for
the relativistic fermions \cite{JJSa}. The same is true in 2+1 dimensional space-time, only in the
presence of the mass term, as the mass-less fermions are spin-less in planar physical
systems \cite{FA2}, like graphene \cite{Gra1,Gra2,Gra3} and topological insulators (TIs)
\cite{TI0,TI}. Massive fermions have been experimentally realized by inducing sub-lattice
density asymmetry in bi-layer graphene under transverse electric field \cite{MM1,GraM02}
and in graphene mono-layer, misaligned with the substrate \cite{GraM2}. The hexagonal
structure of TIs leads to a single Dirac point, with locally inducible gap through
magnetic induction \cite{TEM,TI,TEM1,TIM1}. This Hamiltonian 
emerges as effective description of the low-energy excitations \cite{Gra2}. In this 
regard, aforementioned non-commutativity of the OAM with the Dirac
Hamiltonian, has led to the notion of lattice spin (LS) \cite{LS}, required to 
define a conserved total angular momentum.
\paragraph*{}It has recently been shown that, though of relativistic origin, the
LS has a physical non-relativistic limit \cite{An}. Further, the essential role
of the fermion mass in the existence of LS in gapped graphene is established
\cite{LS}. However, a first-principle derivation of the LS in terms of fundamental
electrons of such planar systems is not given yet. This is crucial in order to identify
the LS in terms of the spin degrees of freedom of constituent 
electrons of the system, which has not been elaborated-upon before.
\paragraph*{}Here, we pursue an ab-initio approach, starting from the 
generalized hopping Hamiltonian in graphene, to physically understand the
origin of the LS, which commutes with the Dirac mass. The presence of two sub-lattices
naturally leads to an emergent spin-symmetry, through individual Heisenberg algebra.
It is shown, through a Schwinger construction of $SU(2)$ algebra,
with fundamental creation/annihilation operators for the sub-lattice electrons, that
it is the SAM, {\it not} OAM of the emergent Dirac fermions, which is directly related to the 
individual spins of the fundamental electrons in graphitic systems. The LS is 
found to be solely arising from the net spin of the system, represented through the 
staggered magnetization of the system, unrelated to the non-local 
nearest neighbor interaction of the fundamental electrons in the tight-binding 
model \cite{Gra2} that correlates the electrons from the two different sub-lattices. 
This non-local hopping term breaks the constructed $SU(2)$ symmetry, as the generators
are no more conserved, and is responsible for massless Dirac dispersion in a hexagonal array.
The latter fact is a further indication that this term is responsible for
non-conservation of the SAM, and induces OAM. The further introduction of a $U(1)$
gauge interaction, owing to the relative sub-lattice displacement \cite{Ando1},
redefines OAM \cite{JJSa} through generalization of linear Dirac momentum in the sense
of covariant derivative. This interaction term, being localized in coordinate space,
cannot effect the LS. 
\paragraph*{}Parity-breaking massive fermions can be realized in graphene and on
the surface of TIs, through the introduction of asymmetry between the electron ground energies 
of the two sub-lattices. This asymmetry is fundamentally related with respective difference of spin-densities,
as the sub-lattices have fixed opposite spins \cite{Neto}. This effectively extends the usual tight-binding Hamiltonian
beyond the hopping term $-t\sum_{\langle i,j\rangle}\left[A^\dagger_iB_j+\text{h.c.}\right]$,
the latter usually expressed modulo the individual sub-lattice ground-state energies, when they are
the {\it same} \cite{Neto}. Including this ground-state contribution, the general expression for
the fundamental electronic Hamiltonian is \cite{LS}, 

\be
H=\sum_i\left[{\cal E}_AA^\dagger_iA_i+{\cal E}_BB^\dagger_iB_i\right]-t\sum_{\langle i,j\rangle}\left[A^\dagger_iB_j+\text{h.c.}\right],\label{02}
\ee
with ${\cal E}_A\neq{\cal E}_B$ for asymmetric sub-lattice vacuum contributions, arising
through lattice deformations \cite{Neto}. This directly leads to non-zero mass
$M=\left({\cal E}_A-{\cal E}_B\right)/2v_F^2$ of the emergent Dirac fermions at low 
energies, with Fermi velocity $v_F$. The electron hopping term has strength $t$, and
$\langle i,j\rangle$ denote nearest neighbor sites at two different sub-lattices $(A,B)$. The respective
fermionic creation $\left(A^\dagger,B^\dagger\right)$ and annihilation $(A,B)$ operators,
anti-commute accordingly, to satisfy the {\it fermionic} Heisenberg algebra \cite{DasPI}:

\be
\{A_i,A_j^\dagger\}=\delta_{ij}=\{B_i,B_j^\dagger\},\label{03}
\ee
with all other anti-commutators vanishing. This allows for the construction of the following
Schwinger's angular momentum algebra \cite{JJS},

\bea
\left[S_+,S_-\right]&=&2S_z,~~~\left[S_z,S_\pm\right]=\pm S_\pm;~~~\text{where}\nonumber\\
S_+&=&\sum_iA^\dagger_iB_i,~~~S_-=\sum_iB^\dagger_iA_i~~~\text{\&}\nonumber\\
S_z&=&\frac{1}{2}\sum_i\left[A^\dagger_iA_i-B^\dagger_iB_i\right].\label{04}
\eea
Here, $i$ denotes a single lattice site, containing two nearest neighbor points
belonging to different sub-lattices.
As the sub-lattices carry opposite electron spins \cite{Neto}, $S_z$ represents the
`net' spin,  or the {\it staggered magnetization} (SM) of graphene electrons, which
is the anti-ferromagnetic order parameter. In the above algebra, the SM plays the role
of projection of the net angular momentum in the diagonal basis. However, the above can
represent the total angular momentum of the system only when the SM commutes with $H$. 
This clearly is {\it not} the case here, in presence of hopping term ($t\neq0$), leading
to,

\bea
\left[H,S_z\right]&=&t\sum_{\langle i,j\rangle}\left[A^\dagger_iB_j-\text{h.c.}\right],\nonumber\\
\left[H,S_\pm\right]&=&\pm 2\left[Mv_F^2S_\pm+tS_z\right].\label{N1}
\eea
From the second relation above, the presence of hopping also prevents the construction
of angular momentum degeneracy in the usual way, through the `ladder operators' $S_\pm$.
Here, $\langle S^2_\pm\rangle=0$, as they anti-commute, which gives rise to a two-fold degeneracy
and thus, represents `spin'.
\paragraph*{}It is to be noted, that the constructed $SU(2)$ algebra comprises of local
operators, which can be resolved in the product vector space $S_A\otimes S_B$, belonging to
($A,B$) sub-lattices. On the other hand, the hopping term entangles the two sub-spaces, 
as it is `non-local', and hence breaks the $SU(2)$ symmetry. The physical implication 
of this symmetry breaking is that the total angular momentum (projection) of the system defined
in Eq. \ref{02} is not completely represented by $S_z$, and an extension is required, as is the case for
Dirac equation \cite{JJSa}. More importantly, this extension must be independent of fundamental
electronic spin, as the same has already been exhausted by the SM.
\paragraph*{}The product vector space of the $SU(2)$ allows for the following matrix formulation
of the generators,

\bea
S_\pm=\sum_i\Phi_i^\dagger\sigma_\pm\Phi_i~~~\text{and}~~~S_z&=&\frac{1}{2}\sum_i\Phi_i^\dagger\sigma_z\Phi_i,\label{05}\\
\text{where}~~~\Phi_i^\dagger&=&\left(A_i^\dagger,B_i^\dagger\right).\nonumber
\eea
This is natural as the Pauli matrices $\sigma_{\pm,z}$ represents $SU(2)$ algebra. In the 
same representation, the Hamiltonian of Eq. \ref{02} can be expressed as,

\be
H=\sum_{\langle i,j\rangle}\Phi_i^\dagger{\cal H}_{ij}\Phi_j,~~~{\cal H}_{ij}=\left(\begin{array}{cc} Mv_F^2\delta_{ij} & -t\\-t & -Mv_F^2\delta_{ij}\end{array} \right),\label{06}
\ee
where the conventional shifting of ${\cal E}_A+{\cal E}_B=0$ has been adopted \cite{LS}.
The above equation clearly shows that in the eigen-basis of $S_z$, only the local term of the
Hamiltonian is diagonal, which depicts the mass of the emergent Dirac fermion. This mass
is of parity-breaking nature, which induces topology in the interacting gauge field as quantum
corrections \cite{TopM1,TopM2,Red,Boy,DM}. From Eqs. \ref{N1}, the mass term is further necessary for the SAM degeneracy.
\paragraph*{}The off-diagonal contribution, responsible for non-conservation of $S_z$, should
then be responsible for the OAM of the emergent Dirac fermions. Indeed, in the momentum ($K$)
space, defined through the Fourier expansion \cite{Gra2},

\be
(A,B)_i=\frac{1}{\sqrt{N}}\sum_j(\alpha,\beta)_j\exp\left(i{\bf R}_i\cdot{\bf Q}_j\right),\label{N01}
\ee
this contribution leads to the linear dispersion in Eq. \ref{01}
at low energies \cite{Gra2}. Here, 

\bea
{\bf R}_i=m_i{\bf a}_1+n_i{\bf a}_2~~~\text{and}~~~{\bf Q}_j&=&r_j{\bf b}_1+s_j{\bf b}_2;\nonumber\\
(m,n,r,s)_i&\in&\mathbb{I},\nonumber
\eea
are lattice vectors in coordinate and momentum spaces, respectively, with corresponding bases 
$\left({\bf a}_{1,2},{\bf b}_{1,2}\right)$. Applying the expansion of Eq. \ref{N01} in
Eq. \ref{06}, with the standard normalization
$\sum_i\exp\left[i{\bf R}_i\cdot\left({\bf Q}_m-{\bf Q}_n\right)\right]=N\delta_{mn}$ \cite{Gra2}, 
the momentum space Hamiltonian matrix can be expressed as,

\be
{\cal H}^{\rm m}_{ij}=Mv_F^2\sigma_z\delta_{ij}-t\sum_{s=-}^+\exp\left(is{\bf R}\cdot{\bf Q}\right)\sigma_{-s}\delta_{ij},\label{N02}
\ee
defined in the basis $\Psi_i^\dagger=\left(\alpha_i^\dagger,\beta_i^\dagger\right)$. Here,
${\bf R}={\bf a}_1+{\bf a}_2$ is the vector separating two nearest neighbor sites in different
sub-lattices and ${\bf Q}$ is the $K$-space vector for the $i$-th site. This Hamiltonian 
satisfies the following commutation relations,

\bea
\left[{\cal H}^{\rm m},S_z\right]&=&-t\sum_{s=-}s\exp\left(is{\bf R}\cdot{\bf Q}\right)S_{-s},\nonumber\\
\left[{\cal H}^{\rm m},S_\pm\right]&=&\pm2Mv_F^2S_\pm\pm2t\exp\left(\pm i{\bf R}\cdot{\bf Q}\right)S_z;\label{N03}\\
{\cal H}^{\rm m}\delta_{ij}&:=&{\cal H}^{\rm m}_{ij},\nonumber
\eea
with the momentum space representations for the $SU(2)$ generators, 

\be
S_\pm=\sigma_\pm,~S_z=\frac{1}{2}\sigma_z,\label{N4}
\ee
ensuring representation of SAM. Eq. \ref{N03} restates that in absence of hopping, the $SU(2)$ symmetry
is exact for the system. For $t\neq0$, deviation from this symmetry is manifested through extensions local
in the momentum space, following non-locality of hopping in the coordinate space, hence yielding a `quantum'
algebra. The relations in Eq. \ref{N03} are valid for graphene at {\it all} energies. 
\paragraph*{}At low energies, near the Dirac points with ${\bf Q}\equiv{\bf K}$, the $K$-space Hamiltonian
of Eq. \ref{N02} reduces to the massive Dirac form \cite{Gra2}:
  
\bea
{\cal H}_D&=&v_F\boldsymbol{\sigma}\cdot{\bf p}+\sigma_zMv_F^2;\label{07}\\
{\bf p}&=&{\bf Q}-{\bf K},~~~{\bf p}\cdot{\bf a}_i\rightarrow0,\nonumber
\eea
which is similar to Eq. \ref{01} with appropriate identifications. This leads to the final 
identification of the operator $S_z$ as the LS, through the result,

\be
[{\cal H}_D,L_z]=-iv_F\left(\boldsymbol{\sigma}\times{\bf k}\right)_z,\label{N2}
\ee
with the `diagonal' projection of the OAM, $L_z={\bf r}\times{\bf p}$, leaving the total
angular momentum projection $J_z=L_z+S_z$ conserved \cite{LS}. Given the mass-spin correspondence 
of the effective theory \cite{Boy}, the LS can be identified as that of the Dirac fermion
$\left(M/2\vert M\vert\right)$, reflected by the corresponding Pauli-Lubanski pseudo-scalar
$W:=k_\mu J^\mu\equiv-Mv_F^2\sigma_z/2$ \cite{DasFT}.
\paragraph*{}Therefore, it is clear that SAM arises from the net spin (staggered magnetization)
of fundamental electrons, whereas the OAM is a result of non-local hopping contribution. The latter
correspondence further makes sense, owing to its `locality' in the $K$-space. In presence of 
relative sub-lattice deformations ${\bf u}({\bf r})$, the resultant $U(1)$ gauge coupling \cite{Ando1} further 
enhances the resolution between SAM and OAM. As a result, the canonical momentum ${\bf p}$ generalizes
to the covariant one $\boldsymbol\Pi={\bf p}-gv_F^2{\bf a}$, with ${\bf a}\propto{\bf u}({\bf r})$
being the effective Abelian gauge field, having coupling strength $g$. Such a system will have
`generalized' OAM \cite{JJSa},

\be
\bar{L}_z={\bf r}\times\boldsymbol\Pi=\epsilon^{ij}r_i\Pi_j.\label{N5}
\ee
The corresponding
expression for planar SAM remains unchanged as \cite{JJSa},

\be
S_z=-\frac{1}{2}\epsilon^{ij}\Sigma_{ij},~~~\Sigma_{ij}=\frac{i}{2}\left[\sigma_i,\sigma_j\right]=-\epsilon_{ijk}\sigma^k,\label{09}
\ee
yielding $S_z=\sigma_z/2$ as before. This is expected as the LS manifests the net fundamental
spin of the hexagonal array, which is independent of the geometry of the system. However, the
orbital part owes to the hopping term which gets effected by relative lattice displacements, which
are interpreted as the $U(1)$ gauge field. Thus, the OAM is changed in the present case.
\paragraph*{}As the LS of the emergent Dirac fermions are shown to be the direct implementation of 
physical spins of the system, it can couple to external electromagnetic fields, and to external
magnetic fields to be specific. However, the spatial dynamics of this emergent fermion is due to
the hopping term, yielding free dispersion. Therefore, provided the long-wavelength approximation 
is still valid to stay confined to the Dirac cone, external magnetic influence can be used for
controlled dynamics of the Dirac fermion. The effective Hamiltonian will have the usual low-energy
form \cite{JJSa},

\be
{\cal H}_{\rm mag}\approx v_F\boldsymbol{\sigma}\cdot{\bf p}+\sigma_zMv_F^2+\frac{g}{M}B_\perp\left(L_z+2S_z\right),\label{10}
\ee
with the last term representing magnetic interaction, with the component $B_\perp$ normal to the 
system plane. This is in conformity to the recently proposed excitonic spin transport in similar
systems \cite{KAPKP}.  
\paragraph*{}In summary, the LS of the emergent Dirac fermion in graphitic systems arises from fundamental
electron spins in the hexagonal lattice. The inherent $SU(2)$ algebra establishes this origin
of the LS, not stated until now. The presence of the gap in the Dirac spectrum is essential 
for the existence of the same, while the hopping term, responsible for the linear dispersion,
breaks the $SU(2)$ symmetry, allowing for an OAM to yield a total angular momentum. This distinction
is further preserved in the presence of $U(1)$ interaction, both emergent and external, the latter
allowing for possible Zeeman coupling.
\paragraph*{Acknowledgement:}PKP would like to thank Dr. Chiranjib Mitra for many useful discussions.


\begin{thebibliography}{99}
\bibitem{JJSa}J. J. Sakurai, {\it Advanced Quantum Mechanics}, (Pearson Education, India, 1967).
\bibitem{FA2}G. V. Dunne, {\it Aspects topologiques de la physique en basse dimension. Topological aspects of low dimensional systems}, (Springer Berlin, Heidelberg, 1999) and references therein.
\bibitem{Gra3} K. Novoselov {\it et al.}, {\it Science} {\bf 306}, 666 (2004).
\bibitem{Gra1} P. R. Wallace, Phys. Rev. {\bf 71}, 622 (1947).
\bibitem{Gra2} G. W. Semenoff, Phys. Rev. Lett. {\bf 53}, 2449 (1984).
\bibitem{TI0}C. L. Kane and E. J. Mele, Phys. Rev. Lett. {\bf 95}, 146802 (2005).
\bibitem{TI}M. Z. Hasan and C. L. Kane, Rev. Mod. Phys. {\bf 82}, 3045 (2010) and references therein.
\bibitem{GraM02}T. Ohta {\it et al.}, Phys. Rev. Lett. {\bf 98}, 206802 (2007); S. Y. Zhou {\it et al.}, {\it Nature Materials} {\bf 6}, 770 (2007); C. R. Woods {\it et al.}, {\it Nature Physics} {\bf 10}, 451 (2013). 
\bibitem{MM1}E. McCann and V. I. Fal'ko, Phys. Rev. Lett. {\bf 96}, 086805 (2006); E. McCann, Phys. Rev. B {\bf 74}, 161403 (2006); T. Ohta {\it et al.}, {\it Science} {\bf 313}, 951 (2006). 
\bibitem{GraM2}B. Hunt {\it et al.}, {\it Science} {\bf 340}, 1427 (2013) and references therein.
\bibitem{TEM}H.-Z. Lu {\it et al.}, Phys. Rev. B {\bf 81}, 115407 (2010͒). 
\bibitem{TEM1}D. Hsieh {\it et al.}, {\it Nature} {\bf 452}, 970 (2008); Y. L. Chen {\it et al.}, {\it Science} {\bf 329}, 659 (2010).
\bibitem{TIM1}Y. L. Chen {\it et al.}, {\it Science} {\bf 329}, 659 (2010); M. I. Katsnelson, F. Guinea and M. A. H. Vozmediano, EPL {\bf 104}, 17001 (2013); O. Vafek and A. Vishwanath, Annu. Rev. of Condens. Matter Phys. {\bf 5}, 83 (2014).
\bibitem{LS}M. Mecklenburg and B. C. Regan, Phys. Rev. Lett. {\bf 106}, 116803 (2011) and references therein.
\bibitem{An}Z. An, F. Q. Liu, Y. Lin and C. Liu, Sci. Rep. {\bf 2}, 388 (2012).
\bibitem{Ando1}K. Ishikawa and T. Ando, J. Phys. Soc. Jpn. {\bf 75}, 84713 (2006).
\bibitem{Neto}A. H. C. Neto {\it et al.},  Rev. Mod. Phys. {\bf 81}, 109 (2009) and references therein.
\bibitem{DasPI}A. Das, {\it Field Theory A Path Integral Approach}, (World Scientific, Singapore, 2006).
\bibitem{JJS}J. J. Sakurai and J. Napolitano, {\it Modern Quantum Mechanics}, (Pearson Education, New York, 2014).
\bibitem{TopM1} J. F. Schonfeld, Nucl. Phys. B {\bf 185}, 157 (1981). 
\bibitem{TopM2} S. Deser, R. Jackiw and S. Templeton, Phys. Rev. Lett. {\bf 48}, 975 (1982); Ann. Phys. (NY) {\bf 140}, 3372 (1982).
\bibitem{Red}A. Redlich, Phys. Rev. Lett. {\bf 52}, 18 (1984); Phys. Rev. D {\bf 29}, 2366 (1984). 
\bibitem{DM}T. W. Appelquist, M. Bowick, D. Karabali and L. C. R. Wijewardhana, Phys. Rev. D {\bf 33}, 3704 (1986).
\bibitem{Boy}D. Boyanovsky, R. Blankenbecler and R. Yahalom, Nucl. phys. B {\bf 270}, 483 (1986).
\bibitem{DasFT} A. Das, {\it Lectures on Quantum Field Theory} (World Scientific, Singapore, 2008)
\bibitem{KAPKP}K. Abhinav and P. K. Panigrahi, arXiv:1504.07955v3.
\end{thebibliography}
\end{document}